\documentclass[12pt]{article}
\usepackage{graphicx}
\usepackage[font=scriptsize]{caption} 
\usepackage{nccmath} 
\usepackage{lineno}
\usepackage{txfonts}

\newcommand\pubdate{\today}

\def\Title#1{\begin{center} {\Large #1 } \end{center}}
\def\Author#1{\begin{center}{ \sc #1} \end{center}}
\def\Address#1{\begin{center}{ \it #1} \end{center}}

\newcommand\pubblock{\rightline{\begin{tabular}{l}  \\ 
         \pubdate  \end{tabular}}}
\newenvironment{Abstract}{\begin{quotation}  }{\end{quotation}}
\newenvironment{Presented}{\begin{quotation} \begin{center} 
             PRESENTED AT\end{center}\bigskip 
      \begin{center}\begin{large}}{\end{large}\end{center} \end{quotation}}

\begin{document}

\begin{titlepage}
 \pubblock
\vfill
\Title{Probing the nature of electroweak symmetry breaking with Higgs boson pairs in ATLAS}
\vfill
\Author{ Bart\l{}omiej \.Zabi\'nski \\
         On behalf of the ATLAS Collaboration}
\Address{The Henryk Niewodnicza\'nski \\ 
         Institute of Nuclear Physics \\ 
           Polish Academy of Sciences}
\vfill
\begin{Abstract}
Constraints on the Higgs boson trilinear self-coupling modifier $\kappa_{\lambda}$ and non-SM HHVV coupling strength $\kappa_{2V}$ are set by combining di-Higgs boson analyses using
$b\bar{b}b\bar{b}$, $b\bar{b}\tau^{+}\tau^{-}$ and $b\bar{b}\gamma\gamma$ decay channels. The data used in these analyses were recorded
by the ATLAS detector at the Large Hadron Collider in proton–proton collisions at $\sqrt{s}$ = 13 TeV and corresponding to an integrated luminosity of 126–139 $fb^{-1}$. The combination of the di-Higgs analyses sets an upper limit of signal strength $\mu_{HH} <$ 2.4 at 95\% confidence level on the di-Higgs production and constraints for $\kappa_{\lambda}$  between -0.6 and 6.6.
The obtained confidence interval for $\kappa_{2V}$ coupling modifier  are [0.1,2.0] . The High Luminosity Large Hadron Collider prospects have been considered as well. The expected signal strength is 0.55 and $\kappa_{\lambda}$ between 0.0 and 2.5 for the baseline scenario.
\end{Abstract}
\vfill
\begin{Presented}
DIS2023: XXX International Workshop on Deep-Inelastic Scattering and
Related Subjects, \\
Michigan State University, USA, 27-31 March 2023 \\
     \includegraphics[width=9cm]{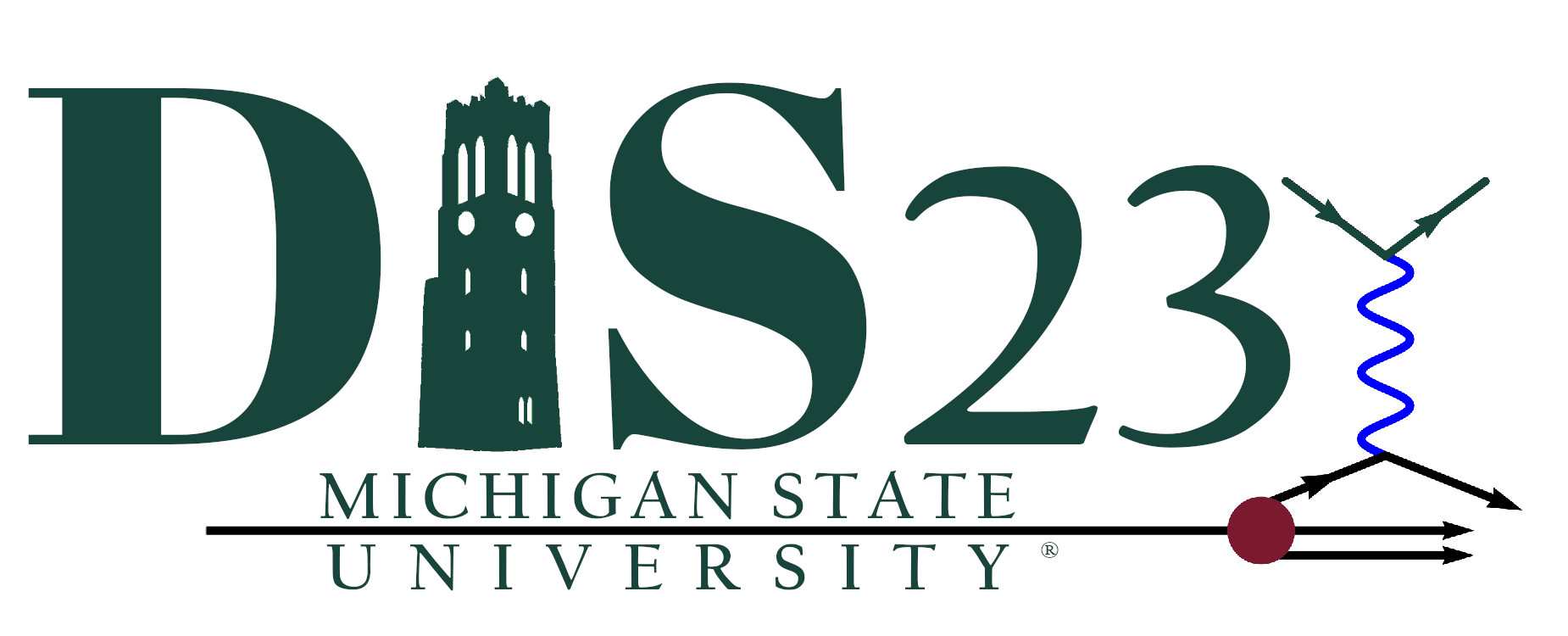}
\end{Presented}
\vfill
\vfill
\vfill

\footnote{$\copyright$ [2023] CERN for the benefit of the ATLAS Collaboration.
Reproduction of this article or parts of it is allowed as specified in the CC-BY-4.0 license}
\end{titlepage}

\section{Introduction}

\begin{figure}[hbp]
  \begin{center}
   \begin{tabular}{@{}cc@{}}
    \includegraphics[width=0.28\textwidth, keepaspectratio]{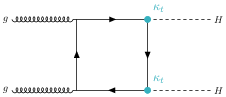} &
    \includegraphics[width=0.28\textwidth, keepaspectratio]{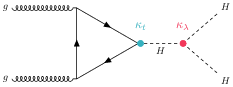} \\
     (a) & (b) \\
     \includegraphics[width=0.22\textwidth, keepaspectratio]{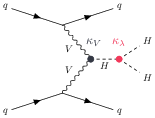} &
     \includegraphics[width=0.22\textwidth, keepaspectratio]{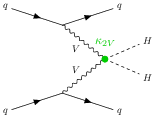} \\
     (c) & (d) \\
    \end{tabular}
  \end{center}
  
 \caption{\label{fig:FeynmanDiag}Examples of leading-order Feynman diagrams for Higgs boson pair production: for ggF production, diagram (a) is proportional to the square of the top-quark Yukawa coupling, while diagram (b) is proportional to the product of the top-quark Yukawa coupling and the Higgs boson self-coupling. For VBF production, diagram (c) and diagram (d) are proportional to the interactions  $\kappa_{V}$ and  $\kappa_{2V}$ between two vectors bosons and two Higgs bosons~\cite{hhcombres}.}

\end{figure}

The discovery of the Higgs boson by ATLAS~\cite{HIGG-2012-27} and CMS~\cite{HIGG_CMS} experiments at the CERN Large Hadron Collider~\cite{artLHC} opened opportunities to measure the boson's properties and test new hypotheses. In the Standard Model~\cite{artSM1,artSM2,artSM3} Higgs boson potential~\cite{Higgp1,Higgp2} provides spontaneous electroweak symmetry breaking that gives rise to its self-interactions. The Higgs boson self-interactions can be described among others by trilinear self-coupling $\lambda_{HHH}$ that can be predicted at lowest order from Fermi constant $G_{F}$ ~\cite{FermiG} and the Higgs mass $m_{H}$ expressed by equation \eqref{eq:1}.
\begin{equation} \label{eq:1}
 \lambda_{HHH} = \frac{m^2_{H}G_{F}}{\sqrt{2}} 
 \end{equation}
 The validity of the SM in the Higgs sector can be tested using the `kappa framework'~\cite{kappf1,kappf2}, in which a coupling modifier $\kappa_{m}$
 is defined as the ratio of the coupling strength between the particle m and Higgs boson to its SM value. The deviation of $\kappa_{m}$ from unity would indicate physics processes beyond SM. For di-Higgs production $\kappa _{\lambda}, \kappa_{t}$, $\kappa_{V}$ and $\kappa_{2V}$ are considered. The $\kappa_{V}$ and  $\kappa_{t}$  describe modifying the SM Higgs boson coupling to W or Z bosons and up-type quarks, respectively. The $\kappa_{2V}$ is related to the VVHH interaction vertex.
The $\kappa _{\lambda}$ coupling modifiers can be measured in the gluon-gluon fusion process (ggF HH) and vector boson fusion (VBF).  
The $\kappa_{V}$ and $\kappa_{2V}$ can be measured in the VBF processes, $\kappa_{2V}$ = 0 in the SM. At the LHC, Higgs boson pair-production is dominated by ggF processes and the overall cross-section in the SM is $\sigma^{SM}_{ggF}(pp\rightarrow HH)$ = $31.05^{+6\%}_{-23\%}$ fb at $\sqrt{s}$ = 13 TeV~\cite{ggFHHpred}. The second largest Higgs pair-production process at LHC is VBF with $\sigma^{SM}_{VBF}$ = $1.72 \pm 0.04$ fb at $\sqrt{s}$ = 13  TeV~\cite{VBF1,VBF2}. Figure \ref{fig:FeynmanDiag} represents leading order diagrams of the ggF and VBF production processes at the LHC featuring coupling modifiers described above.
 
\section{Higgs boson pair production in ATLAS}

\begin{figure}[htbp]
  \begin{center}
    \includegraphics[width=0.4\textwidth, keepaspectratio]{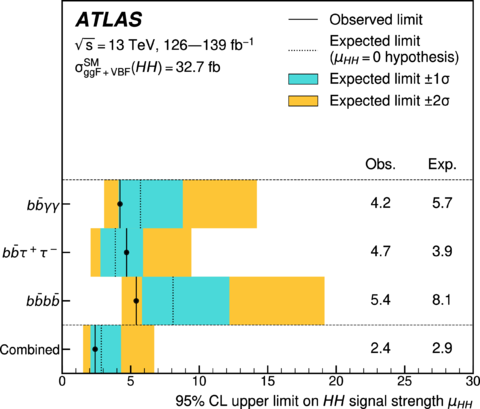}
  \end{center}
  \caption{\label{fig:muCombHH}Observed and expected 95\% CL upper limits on the signal strength for di-Higgs production from the
$b\bar{b}b\bar{b}$, $b\bar{b}\tau^{+}\tau^{-}$ and $b\bar{b}\gamma\gamma$ decay channels, and their statistical combination. The value $m_{H}$ = 125.09 GeV is assumed when deriving the predicted SM cross-section. The expected limit and the corresponding error bands are derived
assuming the absence of the HH process and with all nuisance parameters profiled to the observed data~\cite{hhcombres}.}
\end{figure}

The Higgs pair-production is measured in different decay channels, but in this note, only the most sensitive channels $b\bar{b}b\bar{b}$,  $b\bar{b}\gamma\gamma$ and $b\bar{b}\tau\tau$~\cite{b4b,bbgg,bbtautau} and their combination~\cite{hhcombres} will be discussed. Analysis-specific details can be found in the above-referenced publications. The first measured parameter of the combined results is signal strength $\mu_{HH}$ defined as a ratio of di-Higgs production cross-section including only ggF and VBF processes to its SM prediction of 32.7 fb~\cite{SMpred1, SMpred2, SMpred3}. The expected upper limit of $\mu_{HH}$ is 2.9. The observed upper limit in the absence of HH production, at 95\% CL, is 2.4. The best-fit value obtained from the fit to the data $\mu_{HH}$=-0.7 $\pm$ 1.3. is compatible with the SM prediction of unity, with $p$-value of 0.2.
The obtained limits of $\mu_{HH}$ for each particular channel entering the combination and the results of the combination itself are shown in Figure~\ref{fig:muCombHH} Figure 1.

\begin{figure}[tbp]
  \begin{center}
   \begin{tabular}{@{}cc@{}}
    \includegraphics[width=0.4\textwidth, keepaspectratio]{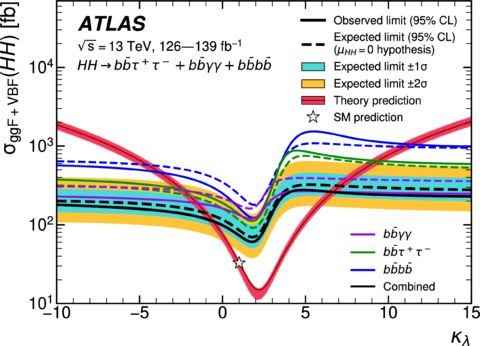} &
    \includegraphics[width=0.4\textwidth, keepaspectratio]{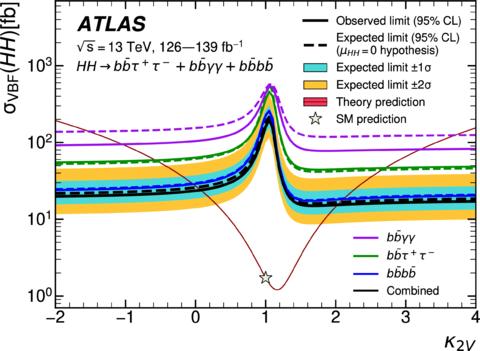} \\
     (a) & (b) \\
    \end{tabular}
  \end{center}
  \caption{\label{fig:vvCombHH}Observed and expected 95\% CL exclusion limits on the production cross-sections of (a) the combined
ggF HH and VBF HH processes as a function of $\kappa_{\lambda}$ and (b) the VBF HH process as a function of $\kappa_{2V}$, for the three
di-Higgs search channels and their combination. The expected limits assume no HH production or no VBF HH
production, respectively. The red line shows (a) the theory prediction for the combined ggF HH and VBF HH cross-section as a function of $\kappa_{\lambda}$ where all parameters and couplings are set to their SM values except for $\kappa_{\lambda}$, and (b) the predicted VBF HH cross-section as a function of $\kappa_{2V}$. The bands surrounding the red cross-section lines indicate the theoretical uncertainty band in (b) is smaller than the width of the plotted line~\cite{hhcombres}.}
\end{figure}

The combination constraints for $\kappa_{\lambda}$ and $\kappa_{2V}$ coupling modifiers, related to $VVHH$ interaction vertex, are shown in Figure~\ref{fig:vvCombHH}. Presented limits for $\kappa_{\lambda}$ and $\kappa_{2V}$ are obtained by using the values of the test statistic as a function of $\kappa_{\lambda}$ in the asymptotic approximation and including the theoretical uncertainty of the cross-section predictions. The obtained confidence interval for $\kappa_{\lambda}$ in each discussed analysis are collected in Table~\ref{tab1}, and the combination constrain at 95\% CL are -0.6 $<\kappa_{\lambda} <$ 6.6 (observed) and -2.1 $<\kappa_{\lambda}< $7.8 (expected).
The $\kappa_{2V}$ coupling modifier for each mentioned channel and their combination have been obtained by fixing all other couplings modifiers to unity and with the expected value derived from the SM hypothesis. The expected and observed at 95\% CL combined constraint for $\kappa_{2V}$ are 0.0 $<\kappa_{2V}<$ 2.1 and 0.1 $<\kappa_{2V}<$ 2.0, respectively.
Table~\ref{tab1} includes $\kappa_{2V}$ constraints for each channel under combination.

\begin{table}
\footnotesize
\begin{center}
\renewcommand{\arraystretch}{1.1}
\begin{tabular}{|c|c|c|c|c|}
\hline
  & $b\bar{b}b\bar{b}$  & $b\bar{b}\tau\tau$ & $b\bar{b}\gamma\gamma$ & combination \\
\hline
Observed 95\% CL & -3.3 $ < \kappa_{\lambda} < $ 11.4 & -2.7 $ < \kappa_{\lambda} < $ 9.5 & -1.4 $ < \kappa_{\lambda} < $ 6.5 & -0.6 $ < \kappa_{\lambda} < $ 6.6\\
Expected 95\% CL &-5.2 $ < \kappa_{\lambda} < $ 11.6 & -3.1 $ < \kappa_{\lambda} < $ 10.2 & -3.2 $ < \kappa_{\lambda} < $ 8.1 &  -2.1 $ < \kappa_{\lambda} < $ 7.8\\
Obs. value & $\kappa_{\lambda} = 6.2^{+3.0}_{-5.2}$ &  $\kappa_{\lambda} = 1.5^{+5.9}_{-2.5}$ & $\kappa_{\lambda} = 2.8^{+2.0}_{-2.2}$ & $\kappa_{\lambda} = 3.1^{+1.9}_{-2.0}$ \\
                    \hline
Observed 95\% CL& 0.0 $ < \kappa_{2V} < $ 2.1   & -0.6 $ < \kappa_{2V} < $ 2.7 & -0.8 $ < \kappa_{2V} < $ 3.0 &  0.1 $ < \kappa_{2V} < $ 2.0 \\
Expected 95\% CL& -0.0 $ < \kappa_{2V} < $ 2.1 & -0.5 $ < \kappa_{2V} < $ 2.7  & -1.6 $ < \kappa_{2V} < $ 3.7 & 0.0 $ < \kappa_{2V} < $ 2.1   
\\
Obs. value & $\kappa_{2V} = 1.0^{+0.7}_{-0.6}$ &  $\kappa_{2V} = 1.5^{+0.7}_{-1.7}$ & $\kappa_{2V} = 1.1^{+1.0}_{-1.0}$ & $\kappa_{2V} = 1.1^{+0.6}_{-0.6}$ \\ 
\hline
\end{tabular}
\caption{\label{tab1}Summary of $\kappa_{\lambda}$ and $\kappa_{2V}$ observed and expected constraints and corresponding observed best-fit values with their uncertainties for the $HH\rightarrow b\bar{b}b\bar{b}$, $HH\rightarrow b\bar{b}\tau\tau$, $HH\rightarrow b\bar{b}\gamma\gamma$, analyses and for the di-Higgs combination. Limits are obtained using the test statistic (-2 ln$\Lambda$) in the asymptotic approximation. The expected constraints are derived under the SM assumption. All other coupling modifiers are fixed to the SM value.
}
\end{center}
\end{table}


%

\section{Prospects of double Higgs production at the HL-LHC}

\begin{figure}[htbp]
  \begin{center}
   \begin{tabular}{@{}cc@{}}
    \includegraphics[width=0.42\textwidth, keepaspectratio]{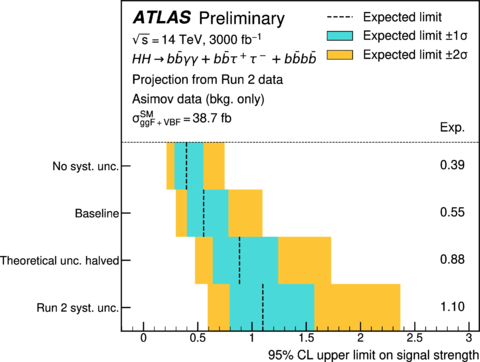} &
    \includegraphics[width=0.42\textwidth, keepaspectratio]{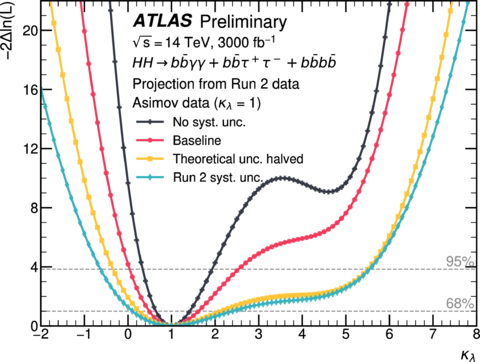} \\
     (a) & (b) \\
    \end{tabular}
  \end{center}
  \caption{\label{fig:CombHHpros} Projected 95\% CL upper limits on the signal strength (a) and negative log-profile-likelihood as a function of $\kappa_{\lambda}$ evaluated on Asimov datasets constructed under the SM hypothesis of $\kappa_{\lambda}$ = 1 (b), combining the $b\bar{b}b\bar{b}$, $b\bar{b}\gamma\gamma$ and $b\bar{b}\tau\tau$ channels at 3000 $fb^{-1}$ and $\sqrt{s}$ = 14 TeV for four uncertainty scenarios described in the text. The limits are derived assuming no HH production.}

\end{figure}

The High Luminosity LHC (HL-LHC) is expected to start in 2029. During the HL-LHC operation it is expected to obtain a total integrated luminosity of 3000 $fb^{-1}$ at a center of mass energy $\sqrt[]{s}$ = 14 TeV. The expected total number of the collected data is significantly larger than the collected data during the LHC Run-2 offers great conditions to measure the Higgs trilinear self-coupling. The prospects of the di-Higgs production at HL-LHC were obtained by scaling full Run-2 distributions and considering multiple scenarios of systematic uncertainties~\cite{HLprospects}. The scale factors are applied to take into account the integrated luminosity and increase of center-of-mass energy. Four cases of future uncertainties are considered:
\begin{itemize}
\item Only statistical uncertainties are considered (No syst. unc).
\item A baseline scenario where the relevant systematic uncertainties were scaled down with respect to expected HL-LHC data delivered~\cite{hllhcpr}. The data-driven background uncertainties are also reduced according to the integrated luminosity. This relies on a 50\% reduction
of the bootstrap uncertainty at the HL-LHC, while the shape uncertainty is assumed to be identical to Run-2.
\item A scenario where experimental uncertainties from Run-2 are unchanged and theoretical uncertainties associated with HH signal are halved (Theoretical unc. halved).
\item The experimental uncertainties from Run-2 are left unchanged (Run-2 syst. unc).
\end{itemize}
Figure~\ref{fig:CombHHpros} represents the result of the prospect studies. The signal strength for the baseline scenario at 95\% CL is 0.55. The constraint on $\kappa_{\lambda}$ in the same scenario is within the range from 0.0 to 2.5. The results for the rest of the scenarios and more details about discussed prospects are in the note~\cite{HLprospects}.


\section{Conclusions}
The double Higgs production searches offer a unique opportunity to investigate the mechanism of electroweak symmetry-breaking. The measurement of $\lambda_{HHH}$ is challenging due to very small cross-sections of the di-Higgs production processes. Constraints are set on the signal strength $\mu$ as well as on $\kappa_{\lambda}$ and $\kappa_{2V}$. Obtained limits scope for each considered channel are wide, but by the combination of all channels, the constraints have been limited to 0.1 $<\kappa_{2V}<$ 2.0 and -0.6 $<\kappa_{\lambda} <$ 6.6. The di-Higgs searches are also a good probe for beyond Standard Model heavy resonance searches~\cite{resbbgg, resbbtautau, res4b}. The ATLAS experiment is in the Run-3 phase, where it is expected to collect two times more data than during Run-2 and finally reach LHC runs with high luminosity. That significant amount of data will offer spectacular results in the mentioned channels and even more, so please stay tuned. \\
\\
{\bf Funding informaion} This work is supported in part by the Polish Ministry of Education and Science project no. 2022/WK/08.

\end{document}